\documentclass[runningheads]{llncs}
\usepackage{graphicx}
\usepackage{booktabs}
\usepackage{float}
\usepackage{verbatimbox}
\usepackage{amsfonts,amssymb}
\usepackage{bbm}
\usepackage{multirow}
\usepackage[misc]{ifsym} 
\usepackage{fontawesome}

\begin{document}
\title{Unsupervised Anomaly Segmentation using Image-Semantic Cycle Translation
}
%

\author{Chenxin Li \and
Yunlong Zhang \and
Jiongcheng Li \and Yue Huang \and Xinghao Ding }

\institute{Xiamen University}

\maketitle              

\begin{abstract}
The goal of unsupervised anomaly segmentation (UAS) is to detect the pixel-level anomalies unseen during training.  
It is a promising field in the medical imaging community,  e.g,  we can use the model trained with only healthy data to segment the lesions of rare diseases.
Existing methods are mainly based on Information Bottleneck, whose underlying principle is modeling the distribution of normal anatomy via learning to compress and recover the healthy data with a low-dimensional manifold, and then detecting lesions as the outlier from this learned distribution.
However, this dimensionality reduction inevitably damages the localization information, which is especially essential for pixel-level anomaly detection.
In this paper, to alleviate this issue, we introduce the semantic space of healthy anatomy in the process of modeling healthy-data distribution.
More precisely, we view the couple of segmentation and synthesis as a special Autoencoder, and propose a novel cycle translation framework with a journey of 'image$\to$semantic$\to$image'.
Experimental results on the BraTS and ISLES databases show that the proposed approach achieves significantly superior performance compared to several prior methods and segments the anomalies more accurately.

\end{abstract}

\section{Introduction}
The segmentation of lesions is common and essential in clinical diagnosis.  The deep learning methods, especially the supervised ones, solve the task automatically and have shown impressive performance.  However, most of them have some limitations, e.g., they need massive training images and the annotations of lesions. The labels are usually expensive to acquire, and even the pathological images are sometimes scarce, especially for rare cases and diseases. Besides, the model suffers the bias toward training data and generalizes poorly to unseen types of lesions.  To meet the challenges, unsupervised anomaly segmentation (UAS) has emerged as a promising technique to detect arbitrary pathologies. In contrast to supervised learning, this method needs only unannotated health images in the training stage, which is more valuable in clinical practice.

Until now, considerable effort has been devoted to UAS in medical imaging analysis, e.g., lesions in brain MRI. Traditional approaches are based on statistical modeling, content-based retrieval , clustering or outlier-detection \cite{taboada2009anomaly}. Recently, many deep learning methods \cite{ae,spae,vae,cae,cevae,bvae,fanogan,gmvae,nguyen2020unsupervised,baur2020steganomaly} have appeared.
The main idea underlying them is to model the distribution of health anatomy with deep representation learning, from which the anomalies can be detected as outliers. Most of these works are based on Information Bottleneck \cite{bottleneck} to learn the normative distribution of healthy data, using Autoencoder, generative adversarial network (GAN), or their siblings. The models learn to compress the healthy samples to a low-dimensional manifold as well as recover them,  and then spot the anomalous structures from the erroneous recoveries during inference.
However, Baur et. al \cite{spae} raised the concern about the loss of localization information during the stage of dimensionality reduction, which is essential for the pixel-level anomaly detection, and combined the Laplacian Pyramid with Autoencoder to alleviate this tissue. Some other works \cite{nguyen2020unsupervised,baur2020steganomaly} introduced the low-level pretext tasks like region-based inpainting or style-transfer between the real and simulated domain in UAS. Nevertheless, at a high level, they still shared the same spirit as the above methods, which we summarize as 'compress and recover'.

Here, our intuition is that a semantic segmentation task, e.g., healthy-tissue segmentation, whose goal is to learn a mapping from image to semantic domain,  usually brings the dimensionality reduction of data and is a natural compress-like process. 
If we further construct another mapping from semantic layout to image domain, these two mappings can be coupled as a special Autoencoder for modeling the healthy-anatomy distribution.
Thus, in this paper, we propose a cycle translation framework, composed of a segmentation module and synthesis module. Once trained,  the representation of healthy data is impliedly expressed in the flow 
of 'image$\to$semantic$\to$image' where the lesions will be the outliers and located from the erroneous recoveries. 

In general, we think the proposed method brings several benefits as following. (a) As an intermediate representation of Information Bottleneck, the semantic layout has a higher resolution than the low-dimensional manifold in previous works, and thus preserves more localization information that is essential for the later accurate anomaly segmentation. (b) The introduced semantic space can encode the objective model of healthy anatomy more explicitly. For example, the probabilistic prediction distributions allocated by a trained classifier are usually statistically discriminative between normal tissues and unseen anomalies, as discussed in Sec \ref{discussion}.
(c) The proposed method can handle supervised healthy-tissue segmentation as well as unsupervised anomaly segmentation simultaneously. Besides, ideally, with the help of any healthy-feature-intention segmentor, e.g., anatomical tissues in brain MRI, the framework has the potential to dispose of arbitrary anomalies. Here, we evaluate the proposed approach for two types of brain lesions on BraTS and ISLES datasets. The experimental results demonstrate that our method surpasses several prior methods and localizes the lesions more accurately.

\section{Method}
\subsection{Overall Concept} \label{method_1}
\begin{figure}[!t] 
    \centering
    \includegraphics[width=0.9\columnwidth,height=6.5cm]{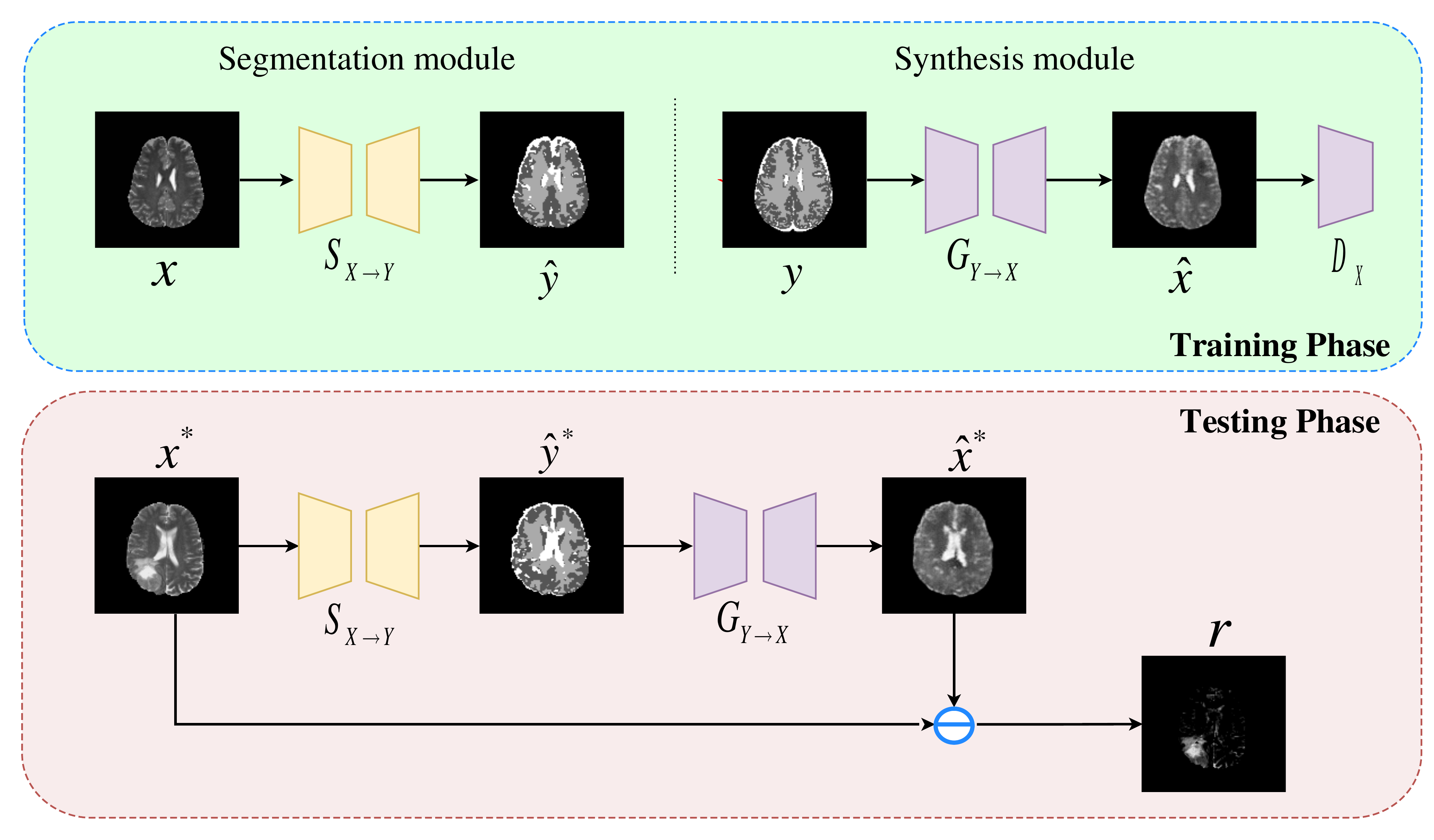}   
    \caption{Pipeline of the proposed image-semantic cycle translation framework, which contains two modules.
    The segmentation module learns a mapping $S$ from the domain of healthy image $X$ to the semantic domain $Y$, e.g., {\itshape the posterior probabilistic distributions of anatomical tissues}, and the synthesis module learns a mapping $G$ from domain $Y$ to domain $X$, associated with an adversarial discriminator $D$.
    During inference, given a query image $x^*$,  we calculate the residuals of the query image and the recovered counterpart $\hat{x}^*$ from the coupled mapping of $S$ and $G$.  For simplicity, the losses involved in optimization are not marked and are described in Sec \ref{method_1} in detail.}
    \label{fig1}  
    \end{figure}

Fig. \ref{fig1} illustrates the pipeline of our proposed framework.  Overall, we model the distribution of healthy anatomy through a cycle translation of 'image-to-semantic' coupled with 'semantic-to-image'
At a high level, the proposed method is characterized by modeling the distribution of healthy data through 'compress and recover', and the semantic layout of healthy tissue can be viewed as the intermediate representation of Information Bottleneck, rather than the low-dimensional manifold in previous works.
Next, we describe the proposed framework in detail. To begin with,
let $x $ be a healthy image and $y\in\{1,2,...,c\}$ be a set of integers representing the pixel-wise labels of anatomical tissues.  

\noindent{\bfseries Segmentation Module}---We utilize a segmentation task for three classes of anatomical tissues in brain MRI: gray matter (GM), white matter (WM), and cerebrospinal fluid (CSF). We train the segmentor S using healthy data with a standard cross-entropy loss, as:
\begin{equation}
\label{eq1}
\mathcal{L}_{SEG}(S) = \mathbb{E}_x[CE(S(x),y)]
\end{equation}
Please note, although the proposed method needs the pixel-wise labels for anatomical tissues, they can be acquired with nearly no pains by some tools like SPM \cite{SPM}.

\noindent{\bfseries Synthesis Module}---We implement the synthesis from semantic space to image domain by a pixel-to-pixel translation with conditional GAN \cite{pix2pix,cgan}, which is composed of a generator G and a discriminator D and trained by an adversarial loss:
\begin{equation}
\label{eq2}
\mathcal{L}_{GAN}(G,D) = \mathbb{E}_x[logD(x)]+\mathbb{E}_y[1-logD(G(y))]
\end{equation}
where G tries to learn to translate the semantic layout to the image-like counterpart while D aims to distinguish between the synthesized image and the real one. 
Besides, a term of L1 loss is introduced to keep consistent image synthesis:
\begin{equation}
\label{eq3}
\mathcal{L}_{1}(G) = \mathbb{E}_{(x,y)}|x-G(y)|
\end{equation}
At last, The module optimized by the combined objective as follows via a minimax two-player game: 
\begin{equation}
\label{eq4}
\mathop {\min }\limits_G\mathop {\max }\limits_D \mathcal{L}_{GAN}(G,D)+\lambda\mathcal{L}_{1}(G)
\end{equation}
where $\lambda$ controls the relative importance of the two objectives.

\noindent{\bfseries Semantic Intermediate Representation}---The semantic representation links the segmentation and synthesis module and plays an important role in the proposed framework. There can be at least two designs for it: continuous semantic representation (probabilistic segmentation maps) or discrete semantic representation (one-hot segmentation maps).
We choose the former based on the intuition that the continuous probabilistic distributions have more powerful modeling capability. We also make the former as the baseline in this paper and conduct the ablation study in Sec \ref{discussion}.  

\subsection{Anomaly Segmentation}
Once the segmentation module and synthesize module are trained, the distribution of healthy data is modeled by the coupled translation of segmentation and synthesis, where we recover the given query image $x^*$.
The regions of erroneous reconstructions are likely to be anomalies, which is the outliers from the learned healthy-data distribution.
Thus, the lesions can be highlighted from the residuals {\bfseries r} between the input image and the recovered image, following most of the prior works:
\begin{equation}
\label{eq6}
r=|x^*-G(S(x^*)|
\end{equation}

\section{Experiments and Results}
\subsubsection{Dataset.}
We evaluated our method on two public datasets: Multimodal Brain Tumor Image Segmentation (BRATS) 2019 \cite{brats1,brats2} and Ischemic Stroke Lesion Segmentation (ISLES) 2015 \cite{isles}, whose lesion types differ with each other. 
The BRATS has 259 GBM (i.e., glioblastoma) and 76 LGG (i.e., lower-grade glioma) volumes, and the ISLES has 28 acute ischemic stroke volumes. We choose 234 GBM volumes of BRATS as the training set
and the remaining volumes as the testing set. 
To guarantee the training set contain only healthy images, we drop the slices where there exist lesions. {\itshape To evaluate the generalization ability among diverse diseases and datasets, we further make the ISLES as another testing set}. 
We make use of only the T2 modality for both two datasets.
Since there are no available labels of anatomical tissues in the training dataset, i.e., BraTS,  we use SPM \cite{SPM}, a MATLAB tool for generating tissue probability maps that represent the average shape of many subjects' brain images, for gray matter (GM), white matter (WM), cerebrospinal fluid (CSF). We use the probability maps for training synthesis module and the one-hot maps for segmentation module.

\subsubsection{Implementation Details.}
We implemented all the methods for comparison as well as ours based on the public code\footnote{https://github.com/StefanDenn3r/Unsupervised-Anomaly-Detection-Brain-MRI}, 
provided by Baur et al. \cite{study},  which ensures a fair comparison.
The input images were down-sampled to the size of 128×128 and normalized to the range of [0, 1]. 
All the experiments were implemented on TensorFlow and an NVIDIA TITAN XP GPU.  
In our method, we use U-Net \cite{Unet} as the segmentation module. 
The backbone of our synthesis module is based on the architecture of the generator and discriminator proposed by Johnson et al. \cite{2016Perceptual}.
We trained the segmentor for 30 epochs, the synthesis module for 15 epochs, using Adam optimizer with learning rate as 2e-4 and $\beta$ as (0.5,0.999). 
The default $\lambda$ in the synthesis module is set as 10 to achieve better visual quality. The batch size was set to 128.

\begin{figure}[]
    \centering
    \includegraphics[width=\textwidth,height=3.5cm]{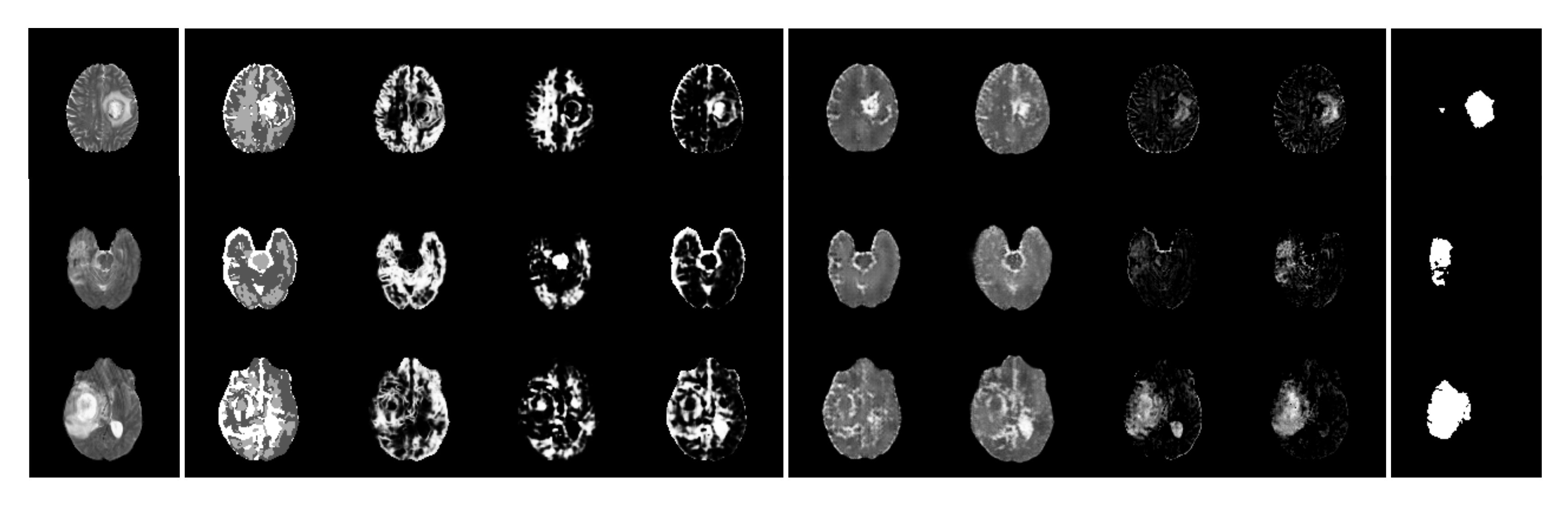} 
    \caption{Visual results on BraTS. 1st column: input slices; 2nd column: discrete prediction maps; 3th$\thicksim$5th column: probabilistic prediction maps for GM, WM and CSF, respectively; 6th$\thicksim$8th column: the reconstructions for our discrete-semantic-intermediated baseline and continuous-semantic-intermediated improved model; 9th column: lesion ground-truth}
    \label{fig2}
    \end{figure} 

\subsubsection{Ablation study}
Comprehensive ablation studies have been performed to compare the different designs of semantic intermediate representation in the proposed framework, i.e., discrete (one-hot segmentation maps) or continuous(probabilistic segmentation maps). We view the former as a baseline and the latter is a improved model as we propose.
As for evaluation metrics for UAS, following Baur et al. \cite{study,spae,baur2020steganomaly}, we utilize Area Under the Precision-Recall curve (AUPRC) and the generally best achievable DICE-score [DICE] as the metrics to assess the performance of UAS. The AUPRC is a common metric to evaluate the comprehensive performance of a classifier under heavy class imbalance, and [DICE] is the theoretical upper-bound for segmentation performance and is determined via a greedy search.

Fig. \ref{fig2} shows the process of the discrete-semantic-intermediated baseline (2nd, 6th, 8th column) and continuous-semantic-intermediated improved model (3rd$\thicksim$5th, 7th, 9th column).
We can observe the recoveries from the improved model have higher visual similarity in healthy regions than those by baseline. It's intuitive since the compulsive discreteness reduces the capacity of semantic intermediate for modeling objective distribution. For example, in the last row, the baseline recovers cerebrospinal fluid poorly and generates the false positives due to its deficiency of modeling norm tissues
As listed in Table \ref{tab1}, the AUPRC and [Dice] achieved by improved model both surpass the baseline, which further indicates the superiority of introducing the continuous semantic intermediate rather than the discrete one in the proposed cycle translation framework.

\begin{table}[]
    \caption{
    The ablation study results of the proposed method on BraTS. 
    }\label{tab1}
    \centering
        \begin{tabular}{lccccc}
         \toprule[1pt]
         \multicolumn{1}{l}{\multirow{2}{*}{~Our Methods~~}}& \multicolumn{2}{c}{~~Semantic Representation} & \multicolumn{2}{c}{Metrics}       \\ \cline{2-5} 
         &~~ Discrete~~      & Continuous~~        & ~~~AUPRC~~          & ~~[DICE]~~           \\ \hline
         ~Proposed Baseline     & $\surd$                 &         & 0.371 & 0.482 \\ \hline
        ~\textbf{Proposed Improved }~  &           & $\surd$                & \textbf{0.511} & \textbf{0.544} \\ 
         \bottomrule[1pt]
        \end{tabular}
    \end{table}

\subsubsection{Comparisons with existing methods}
\begin{figure}[!t]
    \centering
    \includegraphics[width=\textwidth,height=6cm]{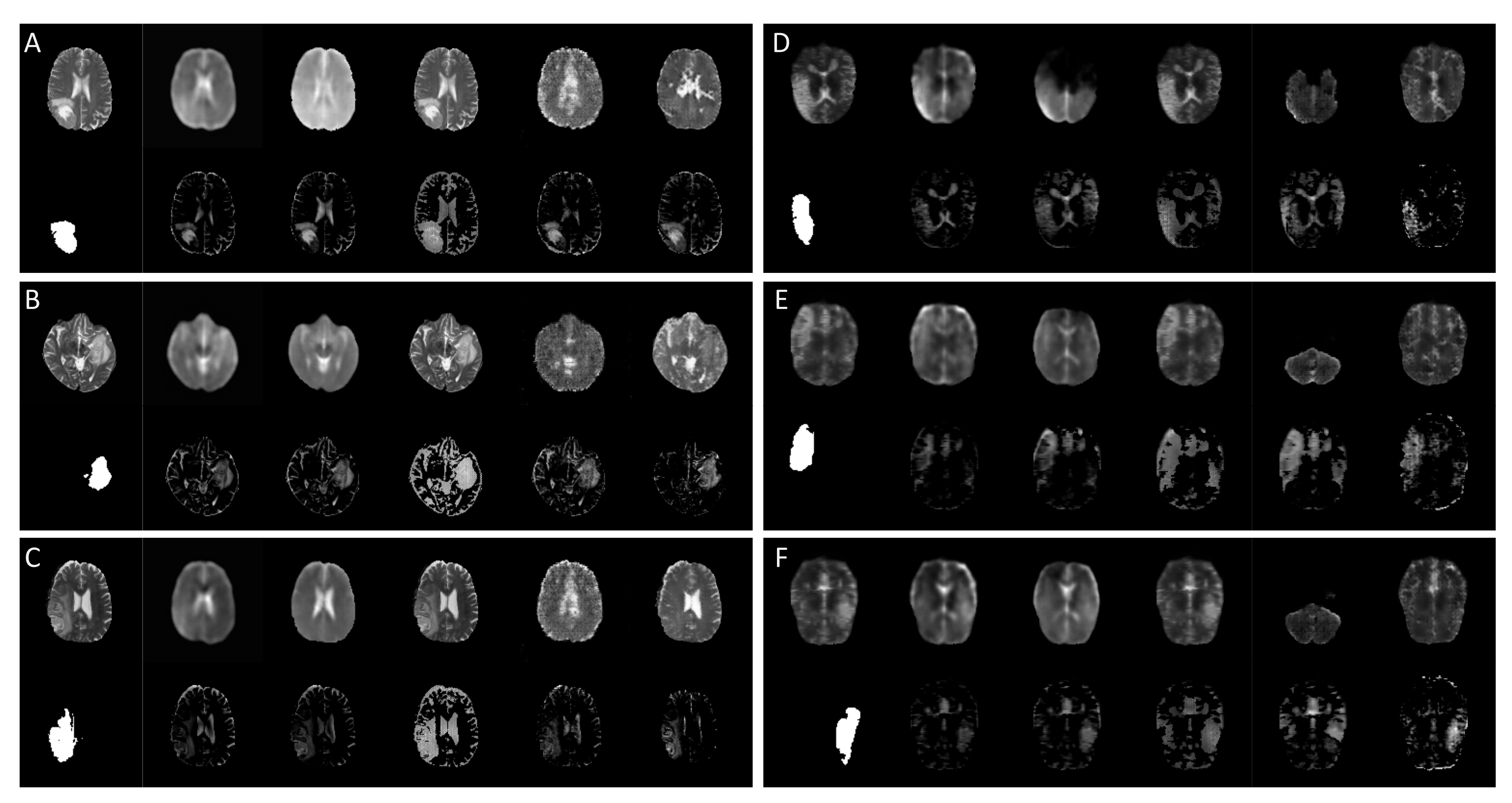}
    \caption{Visual results of method comparisons. A$\thicksim$C: the blocks of results on BraTS; D$\thicksim$F: the blocks of results on ISLES. In each block, 1st row represents the input image, the reconstructions from AE, VAE, GMVAE, fAoGAN and our proposed method; 2nd row represents the lesion ground-truth, the residuals from AE, VAE, GMVAE, fAoGAN and our proposed method.} \label{fig3} 
    \end{figure}

We compare our approach with the existing methods for UAS, including AE \cite{ae}, VAE \cite{vae}, GMVAE \cite{gmvae} and fAnoGAN \cite{fanogan}, on two datasets, BraTS and ISLES.  Table \ref{tab2} and Figure \ref{fig3} presents the quantitative and visual results.
The first three rows of Figure \ref{fig3} corresponds to the results on BraTS.  Compared to other methods, the proposed approach can provide more obvious contrast information between the lesions and healthy tissues, which highlights the lesions and improves anomaly detection.
The reasons come from the following two aspects. On the one hand, the healthy anatomy is recovered successfully.  The proposed approach has high reconstruction fidelity in the regions of healthy tissues, where AE and VAE can generate only blurry reconstructions. fAnoGAN can reconstruct more obviously in those healthy regions, but the visual similarity of normal tissues is still poor.
This makes some highlighted norm tissues in images, like WM and CSF, are also detected, i.e., false positives.
It also confirms the concerns about the loss of localization information in the compression to a low-dimensional manifold. 
On the other hand, the most anomalous regions are recovered poorly by the proposed method, as we are willing to see. Instead, some methods, such as GMVAE, generates nearly the same images as origin input, which ruins the contrast relationship between the anomaly and healthy anatomy, which brings many false negatives and the drop of AUPRC, as shown in Table \ref{tab2}.
In the end, the proposed approach achieves the AUPRC of 0.511 and the [DICE] of 0.544, which is the state-of-the-art performance and outperforms other existing methods by a large margin. 

\begin{table}[!t]
    \caption{
    Quantitative results of our proposed method against prior methods.
    }\label{tab2}
    \centering
	\begin{tabular}{cccccccc}
		\toprule[1pt]
		\multicolumn{1}{c}{\multirow{2}{*}{~~Databases~~}} & \multicolumn{1}{c}{\multirow{2}{*}{~~~~~~Metrics~~~~~~}} & \multicolumn{5}{c}{~~Methods~~}               \\ \cline{3-7} 
		\multicolumn{1}{c}{}                          & \multicolumn{1}{c}{}                        & ~~AE~~     & ~~VAE~~   & ~~GMVAE~~ & ~~fAnoGAN~~   & ~~Ours~~ \\ \hline \hline
		\multirow{2}{*}{BraTS}                        & AUPRC                                       & 0.229  & 0.331  & 0.253   & 0.373  &  \textbf{0.511}     \\
		& {[}DICE{]}                                  & 0.378 & 0.440 & 0.408 & 0.453 & \textbf{0.544}     \\ \hline
		\multirow{2}{*}{ISLES}                        & AUPRC                                       & 0.043  & 0.050   & 0.057   & 0.076  & \textbf{0.110}      \\
		& {[}DICE{]}                                  & 0.112 & 0.117 & 0.116  & 0.164 & \textbf{0.178}      \\ 
		\bottomrule[1pt]
	\end{tabular}
\end{table}

The last three rows of Figure \ref{fig3} are the results on ISLES. All the methods are trained on BraTS and tested on ISLES. This causes the {\itshape domain shift} and makes the performances of whatever methods degrade significantly. More importantly,  {\itshape the lesions in ISLES are more diverse in quantity, pixel intensity, shape (e.g., sometimes very small)}, which also sharply increases the difficulty of this task.  As shown in Table \ref{tab2}, the above disadvantages make the obvious performance degradation compared to that on BraTS. However,  the proposed approach can still achieve the best performance.  For the examples shown in Figure \ref{fig3}, other methods generate massive false positives, but our proposed method avoids most of them still hold on the majority of the effect of anomaly segmentation, which indicates the strong modeling ability and generalization of our method.

\begin{figure}[!t]
    \centering
    \includegraphics[width=0.7\textwidth,height=4.5cm]{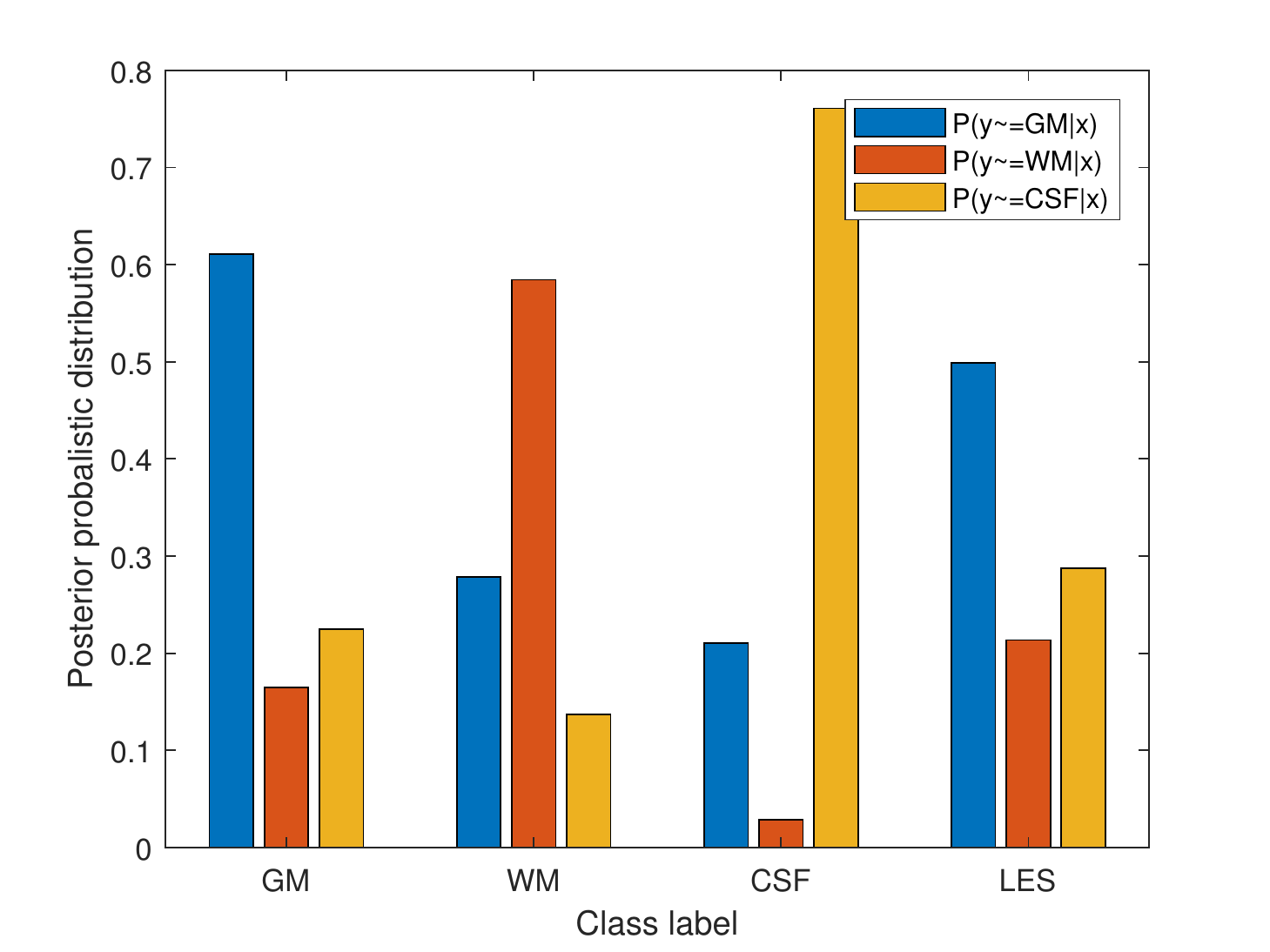}
    \caption{Posterior probabilistic distributions of gray matter (GM), white matter (WM), cerebrospinal fluid (CSF) and lesion (LES), via the trained segmentation module.  Please note: the distributions for a class are acquired by the mean of those of that-class pixels; LES pixels are not counted inside the classes of anatomical tissues.} \label{fig4} 
    \end{figure}

\subsubsection{Discussion on effect of semantic intermediate in brain lesion} \label{discussion}
Figure \ref{fig4} shows the respective posterior probabilistic distributions for different classes,  i.e., GM, WM, CSF, and LES. From two perspectives, we can understand why the proposed semantic intermediate in Information Bottleneck benefits the model capacity of healthy anatomy. On the one hand, this semantic space is explicitly discriminative for different classes by the prediction distributions.  For example, LES has a prediction distribution distinct from most normal tissues.  This discriminative nature benefits the outlier detection.
On the other hand, some normal tissues, like GM, may have a distribution relatively close to LES than other normal anatomies. So in the inference time, some lesion regions may be synthesized as GM.
For T2 modality, nearly the lesions and CSFs are highlighted. So the residuals of these GM-like synthesized regions highlights lesions in turn.

\section{Conclusion}
In this paper, we introduce the semantic space of healthy anatomy in the process of modeling healthy-data distribution.
We view the couple of segmentation and synthesis as a special Autoencoder and propose a novel cycle translation framework with a journey of 'image$\to$semantic$\to$image'.
Experimental results on the BraTS and ISLES databases show that the proposed approach achieves significantly superior performance compared to several prior methods and segments the anomalies more accurately.
 
\newpage 
\bibliographystyle{splncs04}
\bibliography{ref}

\end{document}